\documentstyle[abe,ijmp,12pt]{article}
\begin{document}
\input mssymb.tex
\pagestyle{empty}
\baselineskip20pt
\font\hugeit=cmti10 scaled \magstep4
\def\today{\ifcase\month\or
  January\or February\or March\or April\or May\or June\or
  July\or August\or September\or October\or November\or December\fi
%%  \space\number\day,
  \ \ \number\year}
\vspace*{-40pt}
\rightline{\bf RIMS-1181}
\vspace*{30pt}
\centerline{\cmssB {\hugeit D\/}=26 and Exact Solution to the Conformal-Gauge }
\vskip10pt
\centerline{\cmssB Two-Dimensional Quantum Gravity}
\vskip50pt
\centerline{
 Mitsuo Abe\foot(*,{E-mail: \abemail})
 }
%\vskip3pt
\centerline{\it Research Institute for Mathematical Sciences,
Kyoto University, Kyoto 606-8502, Japan}
\vskip3pt
\centerline{ and }
\vskip3pt
\centerline{
 Noboru Nakanishi\foot({\dagger},{Professor Emeritus of Kyoto University. 
 E-mail: \nnmail})}
%\vskip3pt
\centerline{\it 12-20 Asahigaoka-cho, Hirakata 573-0026, Japan}
\vskip20pt
\centerline{ \today }
%% January 1997}
\vskip100pt
%
%%%%%%%%%%%%%%%%%%%%%% ABSTRACT %%%%%%%%%%%%%%%%%%%%%%%
%
\centerline{\bf Abstract}
The conformal-gauge two-dimensional quantum gravity is formulated in the 
framework of the BRS quantization and solved completely in the Heisenberg
picture: All $n$-point Wightman functions are explicitly obtained.
The field-equation anomaly is shown to exist as in other gauges, but there
is no other subtlety.  At the critical dimension $D=26$ of the bosonic
string, the field-equation anomaly is shown to be absent.
However, this result is not equivalent to the statement that the conformal
anomaly is proportional to $D-26$.   The existence of the FP-ghost number 
current anomaly is seen to be an illusion.
\vfill\eject

\pagestyle{plain}
\setlength{\oddsidemargin}{.5truecm}
\setlength{\textheight}{23.cm}  %{8.85in}
\setlength{\textwidth}{16.cm}
\setlength{\topmargin}{-.5cm}
\setlength{\baselineskip}{19.8pt}
\setlength{\parindent}{25pt}
\textfont0=\tenrm  \textfont1=\teni \textfont2=\tensy \textfont3=\tenex
\def\rm{\fam0 \tenrm} \def\mit{\fam1 } \def\cal{\fam2 }
\def\bf{\tenbf}  \def\it{\tenit} \def\sl{\tensl}
\scriptfont0=\sixrm  \scriptfont1=\sixi  \scriptfont2=\sixsy
\scriptscriptfont0=\smallr \scriptscriptfont1=\smalli
                           \scriptscriptfont2=\smallsy

\rm
%
%%%%%%%%%%%%%%%%%%%%%%%%%%%  Section 1  %%%%%%%%%%%%%%%%%%%%%%%%%%%
%
\Sec{Introduction}
The theory of the bosonic string of a finite length can be consistently
formulated only at the critical dimension $D=26$.\cite{Kato-Ogawa}
On the other hand, the two-dimensional quantum gravity coupled with
$D$ scalar fields can be regarded as the string theory in the $D$-dimensional
spacetime, but in this case the string is not of finite length.
In the conformal gauge, the conformal anomaly proportional to $D-26$ is
obtained in the functional-integral formalism by Fujikawa's method 
based on the functional-integral measure.\cite{Fujikawa}
For covariant gauges (de Donder gauge and some other gauge fixings 
involving differential operator), however, perturbative approach only 
has been available to deduce the conformal anomaly: The two-point
function of the ``energy-momentum tensor'' exhibits a nonlocal term
proportional to $D-26$, which is identified with the conformal 
anomaly.$^{\hbox{\sc \citen{Dusedau}--\citen{Freedman-Lattore-Pilch}}}$
%\cite{Dusedau, Baulieu-Bilal, Rebhan-Kraemmer, Kraemmer-Rebhan, 
%Lattore, Freedman-Lattore-Pilch}
Such a term is obtained also in the perturbative approach to the 
conformal-gauge case.
\par
In our previous paper,\cite{AN1} we have thoroughly reexamined the 
derivations of the conformal anomaly in the covariant gauges.  
Our conclusions are as follows. The proper framework of the two-dimensional
quantum gravity formulated in the Heisenberg picture has no anomaly
for any particular symmetry.  Instead, it has a new-type anomaly, called
``field-equation anomaly'',\cite{AN2} whose existence is confirmed also
in some other two-dimensional models.\cite{AN3,AN4}
By making use of the field-equation anomaly, one {\it can\/} encounter
an anomaly for any particular symmetry such as the conformal anomaly
{\it at one's will\/}.  Especially, the conformal anomaly proportional
to $D-26$ is shown to be obtained by employing a particular perturbative
approach based on the conventional choice of the B-field, but of course
this result has no intrinsic meaning in the Heisenberg picture.
\par
The purpose of the present paper is to extend the consideration made in
Ref.\citen{AN1} to the case of conformal gauge.  This is of particular
interest because in the conformal gauge the conformal anomaly is shown
to be proportional to $D-26$ not only perturbatively but also 
nonperturbatively as stated above.  What are found in the present paper
are as follows.  In the conformal gauge, we can explicitly construct
all $n$-point Wightman functions, which are consistent with the BRS
invariance and the FP-ghost number conservation.  The field-equation
anomaly is shown to exist, and it is proportional to $D-26$ though
it is not equivalent to the conformal anomaly.  There is no such
ambiguity of the critical dimension as was found in the covariant-gauge
case.  The perturbative approach is shown to be inadequate, but it happens
to yield the same value owing to the speciality of the conformal gauge.
\par
The present paper is organized as follows.  In Sec.\ 2, we present
the BRS formulation of the two-dimensional quantum gravity in the 
conformal gauge.  In Sec.\ 3, we show that the theory becomes much 
transparent by rewriting traceless symmetric tensors into vector-like
quantities.  In Sec.\ 4, it is shown that further simplification is 
achieved by introducing light-cone coordinates.  In Sec.\ 5, all $n$-point
Wightman functions are explicitly constructed.  In Sec.\ 6, the existence
of the field-equation anomaly is demonstrated.  In Sec.\ 7, the conformal
anomaly is considered and its connection with the field-equation anomaly
is discussed.  The final section is devoted to the discussion.
\vskip50pt
%
%%%%%%%%%%%%%%%%%%%%%%%%%%%%%  Section 2  %%%%%%%%%%%%%%%%%%%%%%%%%%%%
%
\Sec{Basic formulation}
We present the BRS formulation of the conformal-gauge two-dimensional quantum
gravity.\cite{Yang}
\par
The contravariant gravitational field $g^{\mu\nu}$ is parametrized as 
\begin{eqnarray}
&& g^{\mu\nu}=e^{-\theta}(\eta^{\mu\nu}+h^{\mu\nu}),
\end{eqnarray}
where $\eta^{\mu\nu}$ denotes the Minkowski metric and $h^{\mu\nu}$ is a 
traceless symmetric tensor:
\begin{eqnarray}
&&\eta_{\mu\nu}h^{\mu\nu}=0.
\end{eqnarray}
Let $g^{-1}\equiv\det{g^{\mu\nu}}$ and 
$\tg^{\mu\nu}\equiv(-g)^{1/2}g^{\mu\nu}$; then
\begin{eqnarray}
&&\tg^{\mu\nu}=(\eta^{\mu\nu}+h^{\mu\nu})(1-\det{h^{\sigma\tau}})^{-1/2}.
\end{eqnarray}
It is important to note that $\det{h^{\sigma\tau}}$ is quadratic 
in $h^{\mu\nu}$.
\par
Let $\gdel$ be the conventional BRS transformation and $c^\mu$ be the FP ghost.
From
\begin{eqnarray}
&&\gdel\,g^{\mu\nu}=
 g^{\mu\sigma}\partial_\sigma c^\nu
+g^{\nu\sigma}\partial_\sigma c^\mu
-\partial_\sigma g^{\mu\nu}\cdot c^\sigma \\
\noalign{\noindent together with}
&&\eta_{\mu\nu}\,\gdel h^{\mu\nu}=0,
\end{eqnarray}
we obtain\foot(a,{The fifth term of {(}2.6{)} is missing in Ref.\citen{Yang}.})
\begin{eqnarray}
\gdel h^{\mu\nu}&=&\partial^\mu c^\nu+\partial^\nu c^\mu
+h^{\mu\sigma}\partial_\sigma c^\nu +h^{\nu\sigma}\partial_\sigma c^\mu
-\partial_\sigma h^{\mu\nu}\cdot c^\sigma \nonumber\\
&&-(\eta^{\mu\nu}+h^{\mu\nu})(\partial_\sigma c^\sigma 
 +h^{\sigma\tau}\partial_\sigma c_\tau), \\
\gdel \theta &=&-(\partial_\sigma c^\sigma 
+h^{\sigma\tau}\partial_\sigma c_\tau+\partial_\sigma\theta\cdot c^\sigma).
\end{eqnarray}
For the FP ghost $c^\mu$ and scalar fields $\phi_M\ (M=0,\,1,\,\cdots,\,D-1)$, 
we have
\begin{eqnarray}
&& \gdel c^\mu=-c^\sigma\partial_\sigma c^\mu, \\
&& \gdel \phi_M=-c^\sigma\partial_\sigma \phi_M.
\end{eqnarray}
Let $\tb_{\mu\nu}$ and $\bar c_{\mu\nu}$ be the B field and the FP antighost, 
respectively; they are both traceless symmetric tensors.  As usual, we have
\begin{eqnarray}
&&\gdel \bar c_{\mu\nu}=i\tb_{\mu\nu},\\
&&\gdel \tb_{\mu\nu}=0.
\end{eqnarray}
\par
Now, the Lagrangian density of the conformal-gauge two-dimensional quantum
gravity is given by
\begin{eqnarray}
&&\lag =\lagS+\lagGF+\lagFP, \\
\noalign{\noindent where}
&&\lagS={1\over2}\tg^{\mu\nu}\partial_\mu\phi_M\cdot\partial_\nu\phi^M, \\
&&\lagGF+\lagFP
   ={1\over2}i\gdel\,(\bar c_{\mu\nu}h^{\mu\nu}) \nonumber\\
&&\hspace*{50pt}
   =-{1\over2}\tb^{\mu\nu}h_{\mu\nu}-{i\over2}\bar c_{\mu\nu}\gdel h^{\mu\nu},
\end{eqnarray}
which does not involve the conformal degree of freedom, $\theta$.
\par
The field equations which follows from \eqno(2,12) are as follows.  First, we
have
\begin{eqnarray}
&&h^{\mu\nu}=0.
\end{eqnarray}
The other field equations can be simplified by using \eqno(2,15); especially,
$\det{h^{\sigma\tau}}$ does not contribute to field equations.  Taking the 
tracelessness of $h^{\mu\nu}$, $\tb_{\mu\nu}$ and $\bar c_{\mu\nu}$ into
account, we have
\begin{eqnarray}
&&-\tb_{\mu\nu}-i[\bar c_{\mu\sigma}\partial_\nu c^\sigma
                 +\bar c_{\nu\sigma}\partial_\mu c^\sigma
                 +\partial_\sigma\bar c_{\mu\nu}\cdot c^\sigma
                 -\eta_{\mu\nu}\bar c_{\sigma\tau}\partial^\sigma c^\tau]
  \nonumber \\
&&\qquad 
  +\partial_\mu\phi_M\cdot\partial_\nu\phi^M
   -{1\over2}\eta_{\mu\nu}\partial_\sigma\phi_M\cdot\partial^\sigma\phi^M=0,\\
&&\partial^\mu\bar c_{\mu\nu}=0,\\
&&\partial^\mu c^\nu+\partial^\nu c^\mu
  -\eta^{\mu\nu}\partial_\sigma c^\sigma=0, \\
&&\square \phi_M=0.
\end{eqnarray}
\par
Next, we carry out the canonical quantization.  The canonical variables are
$c^\mu$ and $\phi_M$.  Their canonical conjugates are
\begin{eqnarray}
&&\pi_{c\,\mu}\equiv{\partial \over \partial(\partial_0 c^\mu)}\lag
  =i\bar c_{\mu0}, \\
&&\pi_\phi{}^M\equiv{\partial\over\partial(\partial_0\phi_M)}\lag
  =\partial_0\phi^M.
\end{eqnarray}
Hence, nontrivial canonical commutation relations are
\begin{eqnarray}
&&\{ i\bar c_{\mu 0}(x),\; c^\lambda(y)\}_{x^0=y^0}=-i\delta_\mu{}^\lambda
  \delta(x^1-y^1), \\
&&[\partial_0\phi_M(x),\;\phi^N(y)]_{x^0=y^0}=-i\delta_M{}^N\delta(x^1-y^1).
\end{eqnarray}
The fields $c^\mu$, $\bar c_{\mu\nu}$ and $\phi_M$ are thus free fields
The two-dimensional commutators are given by
\begin{eqnarray}
&& \{ \bar c_{\mu\nu}(x),\; c^\lambda(y)\}=
(\delta_\mu{}^\lambda\partial_\nu
 +\delta_\nu{}^\lambda\partial_\mu
 -\eta_{\mu\nu}\partial^\lambda)D(x-y), \\
&&[\phi_M(x),\;\phi^N(y)]=i\delta_M{}^ND(x-y),\\
\noalign{\noindent where}
&& \qquad D(x)\equiv -{1\over2}\ep(x^0)\theta(x^2).
\end{eqnarray}
\vskip50pt
%
%%%%%%%%%%%%%%%%%%%%%%%%%%  Section 3  %%%%%%%%%%%%%%%%%%%%%%%%%
%
\Sec{Rewriting traceless symmetric tensors}
We encounter three traceless symmetric tensors $h^{\mu\nu}$, $\tb_{\mu\nu}$
and $\bar c_{\mu\nu}$, but their treatment is rather inconvenient because of
their tracelessness.  It is more convenient to rewrite them as if they were
vectors.
\par
Generically, let $X_{\mu\nu}$ be a traceless symmetric tensor:
\begin{eqnarray}
&&X_{\mu\nu}=X_{\nu\mu},\qquad \eta^{\mu\nu}X_{\mu\nu}=0.
\end{eqnarray}
Then it has only two independent components $X_{00}=X_{11}$ and 
$X_{01}=X_{10}$.  We introduce a constant, totally symmetric rank-3 tensorlike
quantity $\xi^{\mu\nu\lambda}$ by
\begin{eqnarray}
\xi^{\mu\nu\lambda}&=&1 \quad \hbox{for}\ \mu+\nu+\lambda\equiv0\ (\hbox{mod.}2)
 \nonumber \\
                   &=&0 \quad \hbox{for}\ \mu+\nu+\lambda\equiv1\ (\hbox{mod.}2).
\end{eqnarray}
Then we define $X^\lambda$ by
\begin{eqnarray}
&&X^\lambda\equiv {1\over\sqrt2}\xi^{\lambda\mu\nu}X_{\mu\nu}.
\end{eqnarray}
We have the following formulae:
\begin{eqnarray}
&&\xi_{\mu\nu\lambda}\xi^{\lambda\sigma\tau}
 =\delta_\mu{}^\sigma\delta_\nu{}^\tau
 +\delta_\mu{}^\tau\delta_\nu{}^\sigma
 -\eta_{\mu\nu}\eta^{\sigma\tau}, \\
&&\xi_{\lambda\mu\nu}\xi^{\mu\nu\rho}=2\delta_\lambda{}^\rho,\\
&&\eta^{\mu\nu}\xi_{\mu\nu\lambda}=0; \\
&&X_{\mu\nu}={1\over\sqrt2}\xi_{\mu\nu\lambda}X^\lambda, \\
&&\sqrt{2}\partial^\nu X_{\mu\nu}=\xi_{\mu\nu\lambda}\partial^\nu X^\lambda,\\
&&\sqrt{2}\xi^{\sigma\tau\mu}\partial^\nu X_{\mu\nu}
=(\delta_\lambda{}^\tau\partial^\sigma
 +\delta_\lambda{}^\sigma\partial^\tau
 -\eta^{\sigma\tau}\partial_\lambda)X^\lambda.
\end{eqnarray}
\par
Applying the above consideration to $h^{\mu\nu}$, $\tb_{\mu\nu}$ 
and $\bar c_{\mu\nu}$, we introduce $h_\lambda$, $\tb^\lambda$ 
and $\bar c^\lambda$.  Then \eqno(2,6), \eqno(2,10) and \eqno(2,11) are
rewritten as
\begin{eqnarray}
&&\gdel h_\lambda=\sqrt2\xi_{\lambda\mu\nu}\partial^\mu c^\nu
   +\xi_{\lambda\mu\nu}\xi^{\mu\sigma\tau}h_\sigma\partial_\tau c^\nu
   -\partial_\nu(h_\lambda c^\nu)
   -{1\over\sqrt2}h_\lambda\xi^{\nu\sigma\tau}h_\nu\partial_\sigma c_\tau, \\
&&\gdel \bar c^\lambda = i\tb^\lambda,\\
&&\gdel \tb^\lambda=0,
\end{eqnarray}
respectively.  Correspondingly, \eqno(2.14) is rewritten as
\begin{eqnarray}
&&\lagGF+\lagFP=-{1\over2}\tb^\lambda h_\lambda 
  -{i\over2}\bar c^\lambda\gdel h_\lambda.
\end{eqnarray}
Field equations \eqno(2,15){,} \eqno(2,16) and \eqno(2,17) become
\begin{eqnarray}
&&h_\mu=0,\\
&&-\tb^\mu-i[\xi_{\sigma\tau\rho}\xi^{\rho\mu\lambda}
   \bar c^\sigma\partial_\lambda c^\tau 
  + \partial_\sigma\bar c^\mu\cdot c^\sigma]
  + {1\over\sqrt2}\xi^{\mu\sigma\tau}\partial_\sigma\phi_M
     \cdot\partial_\tau\phi^M =0, \\
&&\xi_{\lambda\mu\nu}\partial^\mu\bar c^\nu=0,
\end{eqnarray}
respectively.  On the other hand, \eqno(2.18) is rewritten as
\begin{eqnarray}
&&\xi_{\lambda\mu\nu}\partial^\mu c^\nu=0.
\end{eqnarray}
From \eqno(3,15){,} with the help of field equations{,} we obtain
\begin{eqnarray}
&&\xi_{\lambda\mu\nu}\partial^\mu \tb^\nu=0.
\end{eqnarray}
Evidently, \eqno(3,18) is the BRS transform of \eqno(3,16).
Furthermore, the identity \eqno(3,6) is rewritten as
\begin{eqnarray}
&&\xi_{\lambda\mu\nu}\partial^\mu x^\nu=0.
\end{eqnarray}
\par
As in the de Donder-gauge quantum Einstein gravity,\cite{N1,NO}
it is natural to introduce ``supercoordinate'' $X^\nu\equiv(x^\nu,\;\tb^\nu,\;
c^\nu,\;\bar c^\nu)$; then \eqno(3,16)-\eqno(3,19) are unified into
\begin{eqnarray}
&&\xi_{\lambda\mu\nu}\partial^\mu X^\nu=0.
\end{eqnarray}
Accordingly, we have many conservation laws:
\begin{eqnarray}
&&\partial^\mu(\xi_{\mu\nu\lambda}X^\nu)=0,\\
&&\partial^\mu(\xi_{\mu\nu\lambda}X^\nu X^\lambda)=0.
\end{eqnarray}
From \eqno(3,21) and \eqno(3,22), we obtain many symmetry generators, 
most of which are spontaneously broken.  Unfortunately, it is very difficult
to find the corresponding symmetry transformations which leave the action 
invariant, because in the action we must not use the field equations, 
especially \eqno(3,14).\cite{AN5}
\par
The two-dimensional anticommutation relation \eqno(2,24) is rewritten as
\begin{eqnarray}
&&\{ c^\rho(x),\;\bar c^\lambda(y)\}
=\sqrt2\xi^{\rho\lambda\nu}\partial_\nu D(x-y).
\end{eqnarray}
The two-dimensional commutators involving $\tb^\mu$ can be calculated 
by expressing $\tb^\mu$ in terms of $\bar c^\rho$, $c^\lambda$ and $\phi_M$
from \eqno(3,15).  For example, we have
\begin{eqnarray}
[\tb^\lambda(x),\;\phi_M(y)]=i\sqrt2\xi^{\lambda\mu\nu}\partial_\mu\phi_M(x)
 \cdot\partial_\nu D(x-y).
\end{eqnarray}
Unfortunately, if $[\tb^\lambda(x),\;\tb^\rho(y)]$ is calculated by this
method or by using the BRS transform, we obtain various expressions which 
apparently look different, depending on the ways of calculation.  
This problem is resolved in next section.
\vskip50pt
%
%%%%%%%%%%%%%%%%%%%%%%%%%%%%%  Section 4  %%%%%%%%%%%%%%%%%%%%%%%%%%
%
\Sec{Use of light-cone coordinates}
Two-dimensional commutators are much simplified if we use light-cone
coordinates $x^\pm\equiv(x^0\pm x^1)/\sqrt2$.
This is because in the light-cone coordinates we have $\xi_{\mu\nu\lambda}=0$
only except
\begin{eqnarray}
&&\xi_{+++}=\xi_{---}=\sqrt2.
\end{eqnarray}
Therefore, \eqno(3,20) reduces to 
\begin{eqnarray}
\partial_\mp X^\pm=0,
\end{eqnarray}
that is, $X^\pm$ is a function of $x^\pm$ only.  
Furthermore, \eqno(2,26) yields
\begin{eqnarray}
&&\partial_\pm D(x)=-{1\over2}\delta(x^\pm).
\end{eqnarray}
Hence we always have
\begin{eqnarray}
&&[X^+(x),\;Y^-(y)]=0
\end{eqnarray}
because there is no mixing of $x^+$ and $y^-$.
Furthermore, since there is a complete symmetry in $+\ \leftrightarrow\ -$,
it is sufficient to calculate $[X^+(x),\;Y^+(y)]$ only.
\par
First, the two-dimensional commutation relations \eqno(3,23) and \eqno(2,25)
are simplified into
\begin{eqnarray}
&&\{c^+(x),\;\bar c^+(y)\}=-\delta(x^+-y^+), \\
&&[\partial_+\phi_M(x),\;\phi^N(y)]=-{i\over2}\delta_M{}^N\delta(x^+-y^+), 
\end{eqnarray}
respectively.  Using
\begin{eqnarray}
&&\tb^+(x)=-i(2\bar c^+\partial_+c^+ 
              +\partial_+\bar c^+\cdot c^+)
           +\partial_+\phi_M\cdot\partial_+\phi^M,
\end{eqnarray}
which follows from \eqno(3,15), we calculate the commutators involving $\tb^+$.
We have
\begin{eqnarray}
[ \tb^+(x),\;c^+(y) ]
 &=&-i[c^+(x)\delta'(x^+-y^+)+2\partial_+c^+(x)\cdot\delta(x^+-y^+)], \\  
{[} \tb^+(x),\;\bar c^+(y) ]
 &=&i[\bar c^+(x)+\bar c^+(y)]\delta'(x^+-y^+)\nonumber\\
 &=&i[2\bar c^+(x)\delta'(x^+-y^+)
      +\partial_+\bar c^+(x)\cdot\delta(x^+-y^+)],\\
{[} \tb^+(x),\;\phi_M(y) ]
 &=&-i\partial_+\phi_M(x)\cdot\delta(x^+-y^+), \\
{[} \tb^+(x),\;\tb^+(y) ]
 &=&i[\tb^+(x)+\tb^+(y)]\delta'(x^+-y^+).
\end{eqnarray}
Evidently, \eqno(4,11) is obtained also as the BRS transform of \eqno(4,9).
\vskip50pt
%
%%%%%%%%%%%%%%%%%%%%%%%%%%%%  Section 5  %%%%%%%%%%%%%%%%%%%%%%%%%%%%%%
%
\Sec{Wightman functions}
Since all two-dimensional commutation relations have explicitly been obtained,
we can calculate all multiple commutators.  Then, according to the prescription
given in our previous papers,\cite{AN7,AN6}$^,$\cite{AN2}$^,$\foot(b,{For a summary, see 
Ref.\citen{AN1}.}) we can construct $n$-point Wightman functions
$\Wightman{\vphi_1(x_1)\cdots\vphi_n(x_n)}$, where $\vphi_j(x)$ denotes 
a generic field.
\par
It is natural to set all 1-point functions equal to zero.
Then, as is seen from multiple commutators, nonvanishing truncated $n$-point
Wightman functions are those which consist of $(n-2)$ $\tb^+$'s and of
$c^+$ and $\bar c^+$ or two $\phi_M$'s.
\par
The nonvanishing 2-point functions are as follows.
From \eqno(2,25) we have
\begin{eqnarray}
&&\Wightman{\phi_M(x_1)\phi^N(x_2)}=\delta_M{}^N\Dp(x_1-x_2),
\end{eqnarray}
where $\Dp(x)$ denotes that positive-energy part of $iD(x)$.
Since 
\begin{eqnarray}
&&\partial_+\Dp(x)=-{1\over4\pi}{1\over x^+-i0},
\end{eqnarray}
\eqno(5,1) implies that
\begin{eqnarray}
&&\partial_+{}^{x_1}\Wightman{\phi_M(x_1)\phi^N(x_2)}
 =-{1\over4\pi}\delta_M{}^N{1\over x_1{}^+-x_2{}^+-i0},
\end{eqnarray}
which is deduced directly from \eqno(4,6).
On the other hand, \eqno(4,5) implies that
\begin{eqnarray}
&&\Wightman{\bar c^+(x_1)c^+(x_2)}=\Wightman{c^+(x_1)\bar c^+(x_2)}
={i\over2\pi}{1\over x_1{}^+-x_2{}^+-i0}.
\end{eqnarray}
\par
We proceed to the 3-point functions.
From the double commutators,
\begin{eqnarray}
&&[\;[\phi_M(x_1),\;\tb^+(x_2)],\;\phi^N(x_3)]
   ={1\over2}\delta_M{}^N\delta(x_1{}^+-x_2{}^+)\delta(x_2{}^+-x_3{}^+),\\
&&\{\;[c^+(x_1),\;\tb^+(x_2)],\;\bar c^+(x_3)\} \nonumber \\
&&\quad
   =i[\delta'(x_1{}^+-x_2{}^+)\delta(x_2{}^+-x_3{}^+)
     -2\delta(x_1{}^+-x_2{}^+)\delta'(x_2{}^+-x_3{}^+)],
\end{eqnarray}
together with $[\;[\phi_M,\;\phi^N],\;\tb^+]=[\;\{c^+,\;\bar c^+\},\;\tb^+]=0$,
we deduce\foot(c,{For Wightman functions of other orderings, $-i0$ is 
changed into $+i0$ appropriately {(}and a minus sign is inserted for 
the exchange of $c$ and $\bar c${)}.})
\begin{eqnarray}
&&\Wightman{\phi_M(x_1)\tb^+(x_2)\phi^N(x_3)}
  =-{1\over2(2\pi)^2}\delta_M{}^N
    {1\over x_1{}^+-x_2{}^+-i0}{1\over x_2{}^+-x_3{}^+-i0}, \hspace*{30pt}\\
&&\Wightman{c^+(x_1)\tb^+(x_2)\bar c^+(x_3)}
  ={i\over(2\pi)^2}
   \bigg[ {1\over (x_1{}^+-x_2{}^+-i0)^2}\cdot{1\over x_2{}^+-x_3{}^+-i0}
   \nonumber \\
&&\hspace*{180pt}
   -2{1\over x_1{}^+-x_2{}^+-i0}\cdot{1\over (x_2{}^+-x_3{}^+-i0)^2} \bigg].
\end{eqnarray}
\par
We extend the above analysis to the $n$-point functions.
As $(n-1)!$ independent $(n-1)$-ple  commutators, we adopt those whose first
member is $\phi_M(x_1)$ or $c^+(x_1)$.
Then those which do not have $\phi^N(x_k)$ or $\bar c^+(x_k)$ as the last
member $(k=n)$ vanish.  
\par
It is easy to prove by mathematical induction that
\begin{eqnarray}
&&[\,\cdots\,[\;[\phi_M(x_1),\;\tb^+(x_2)],\;\tb^+(x_3)],\;
   \cdots,\;\phi^N(x_n)]  \nonumber \\
&&\ \  =-{1\over2}\delta_M{}^Ni^{n-1}
  \Rpartial2\,\cdots\,\Rpartial{n-2}
   [\delta(x_1{}^+-x_2{}^+)\cdots\delta(x_{n-1}{}^+-x_n{}^+)]
  \ \  (n\geqq3),  \hspace*{30pt}
\end{eqnarray}
where the superscript R of $\Rpartial{k}$ means that $\partial_k$ acts 
only on the right-hand factor among the ones involving $x_k{}^+$; for instance,
\begin{eqnarray}
&&\Rpartial2[\delta(x_1{}^+-x_2{}^+)\delta(x_2{}^+-x_3{}^+)
             \delta(x_3{}^+-x_4{}^+)] \nonumber \\
&&\qquad
  =\delta(x_1{}^+-x_2{}^+)\delta'(x_2{}^+-x_3{}^+)\delta(x_3{}^+-x_4{}^+).
\end{eqnarray}
From \eqno(5,9), we deduce
\begin{eqnarray}
&&\Wightman{\phi_M(x_1)\tb^+(x_2)\cdots\tb^+(x_{n-1})\phi^N(x_n)} \nonumber\\
&&\quad
 =-{1\over2(2\pi)^{n-1}}\delta_M{}^N
   \sum_{P(j_2,\,\cdots,\,j_{n-1})}^{(n-2)!}
   \Rpartial{j_2}\cdots\Rpartial{j_{n-2}}
   \bigg[{1\over x_1{}^+-x_{j_2}{}^+-i0}{1\over x_{j_2}{}^+-x_{j_3}{}^+\mp i0}
   \nonumber \\
&&\hspace*{100pt}
    \cdots {1\over x_{j_{n-2}}{}^+-x_{j_{n-1}}{}^+ \mp i0}
           {1\over x_{j_{n-1}}{}^+-x_{n}{}^+ -i0} \bigg] \quad
   (n\geqq3), \hspace*{30pt}
\end{eqnarray}
where $P(j_2,\,\cdots,\,j_{n-1})$ stands for a permutation of $(2,\,3,\,\cdots,
\, n-1)$, and 
\begin{eqnarray}
x_j{}^+-x_k{}^+\mp i0&=&x_j{}^+-x_k{}^+-i0 \ \ \hbox{for}\ j<k \nonumber \\
                     &=&x_j{}^+-x_k{}^++i0 \ \ \hbox{for}\ j>k.
\end{eqnarray}
\par
We proceed to the $n$-point function involving $c^+$ and $\bar c^+$.
First, we rewrite \eqno(4,8) as
\begin{eqnarray}
&&[c^+(x_1),\;\tb^+(x_2)]=i(\Lpartial2+2\Rpartial2)
  [\delta(x_1{}^+-x_2{}^+)c^+(x_2)],
\end{eqnarray}
where $\Lpartial{k}$ acts only on the left-hand factor among the ones involving
$x_k{}^+$.
By making use of \eqno(5,13), it is easy to show that
\begin{eqnarray}
&&\{\,\cdots[\;[c^+(x_1),\;\tb^+(x_2)],\;\tb^+(x_3)],
    \;\cdots,\;\bar c^+(x_n)\} \nonumber \\
&&\qquad =-i^{n-2}(\Lpartial2+2\Rpartial2)
            \cdots(\Lpartial{n-1}+2\Rpartial{n-1})
          [\delta(x_1{}^+-x_2{}^+)\cdots\delta(x_{n-1}{}^+-x_n{}^+)]. 
  \nonumber \\
&& % \hspace*{50pt}
\end{eqnarray}
Hence, as above, we have
\begin{eqnarray}
&&\hspace*{-15pt}
\Wightman{c^+(x_1)\tb^+(x_2)\cdots\tb^+(x_{n-1})\bar c^+(x_n)} \nonumber \\
&&={i\over (2\pi)^{n-1}}\sum_{P(j_2,\,\cdots,\,j_{n-1})}^{(n-2)!}
  (\Lpartial{j_2}+2\Rpartial{j_2})
   \cdots(\Lpartial{j_{n-1}}+2\Rpartial{j_{n-1}})   \nonumber \\
&&\ \ 
 \bigg[ {1\over x_1{}^+-x_{j_2}{}^+-i0}{1\over x_{j_2}{}^+-x_{j_3}{}^+\mp i0}
       \cdots {1\over x_{j_{n-2}}{}^+-x_{j_{n-1}}{}^+\mp i0}
              {1\over x_{j_{n-1}}{}^+-x_n{}^+- i0} \bigg]. %\hspace*{40pt}
  \nonumber \\
&&
\end{eqnarray}
\par
Of course, the completely analogous formulae hold for 
$\Wightman{\phi_M\tb^-\cdots\tb^-\phi^N}$ and 
$\Wightman{c^-\tb^-\cdots\tb^-\bar c^-}$.
All mixed Wightman functions vanish.
\par
If the BRS invariance is not broken, the following Ward-Takahashi identities
must hold:
\begin{eqnarray}
0&=&\Wightman{\gdel[\phi_M(x_1)\tb^+(x_2)\cdots\tb^+(x_{n-2})\bar c^+(x_{n-1})
    \phi^N(x_n)]} \nonumber \\
 &=&i\Wightman{\phi_M(x_1)\tb^+(x_2)\cdots\tb(x_{n-1})\phi^N(x_n)} \nonumber\\
  &&-\Wightman{c^+(x_1)\partial_+\phi_M(x_1)\cdot\tb^+(x_2)\cdots\tb^+(x_{n-2})
      \bar c^+(x_{n-1})\phi^N(x_n)} \nonumber \\
  &&+\Wightman{\phi_M(x_1)\tb^+(x_2)\cdots\tb^+(x_{n-2})\bar c^+(x_{n-1})
      c^+(x_n)\partial_+\phi^N(x_n)}, \\
0&=&\Wightman{\gdel[c^+(x_1)\tb^+(x_2)\cdots\tb^+(x_{n-2})
              \bar c^+(x_{n-1})\bar c^+(x_n)]} \nonumber \\
 &=&-i\Wightman{c^+(x_1)\tb^+(x_2)\cdots\tb^+(x_{n-2})\tb^+(x_{n-1})
                \bar c^+(x_n)} \nonumber \\
  &&+i\Wightman{c^+(x_1)\tb^+(x_2)\cdots\tb^+(x_{n-2})\bar c^+(x_{n-1})
                \tb^+(x_n)} \nonumber \\
  &&-\Wightman{c^+(x_1)\partial_+c^+(x_1)\cdot\tb^+(x_2)\cdots\tb^+(x_{n-2})
                \bar c^+(x_{n-1})\bar c^+(x_n)}.
\end{eqnarray}
We have explicitly confirmed for $n=3,\,4$ that \eqno(5,11) and \eqno(5,15) 
are indeed consistent with \eqno(5,16) and \eqno(5,17).
For example, \eqno(5,16) for $n=3$ becomes as follows:
\begin{eqnarray}
&&i\Wightman{\phi_M(x_1)\tb^+(x_2)\phi^N(x_3)} \nonumber \\
&&-\Wightman{c^+(x_1)\bar c^+(x_2)}\Wightman{\partial_+\phi_M(x_1)\phi^N(x_3)}
   \nonumber \\
&&+\Wightman{\phi_M(x_1)\partial_+\phi^N(x_3)}\Wightman{\bar c^+(x_2) c^+(x_3)}
   \nonumber \\
&&\quad =-{i\over 2(2\pi)^2}\delta_M{}^N{1\over x_1{}^+-x_2{}^+-i0}
                            {1\over x_2{}^+-x_3{}^+-i0} \nonumber \\
&&\qquad -{i\over 2\pi}\cdot{1\over x_1{}^+-x_2{}^+-i0}\cdot
          \left(-{1\over 4\pi}\right)\delta_M{}^N{1\over x_1{}^+-x_3{}^+-i0}
          \nonumber \\
&&\qquad +{1\over4\pi}\delta_M{}^N{1\over x_1{}^+-x_3{}^+-i0}
          \cdot{i\over2\pi}{1\over x_2{}^+-x_3{}^+-i0} \nonumber\\
&&\quad =0.
\end{eqnarray}
The confirmation of \eqno(5,17) for $n=4$ needs rather long calculation.
\vskip50pt
%
%%%%%%%%%%%%%%%%%%%%%%%%%%%%  Section 6  %%%%%%%%%%%%%%%%%%%%%%%%%%%%
%
\Sec{Field-equation anomaly}
The existence of the field-equation anomaly was found in the various
massless two-dimensional models$^{\hbox{\sc \citen{AN2}--\citen{AN4}}}$:
%\cite{AN2,AN3,AN4}: 
One of field equations is broken at the level of the representation 
in terms of state vectors, {\it modulo\/} a field equation which is 
obtained from the original field equation by differentiating it once
or twice but has the same degree of freedom as its.
In this section we discuss the field-equation anomaly in the conformal-gauge
two-dimensional quantum gravity.
\par
We write \eqno(2,16) as
\begin{eqnarray}
&&2{\delta\over\delta h^{\mu\nu}}\int d^2x\,\lag
\equiv \calT_{\mu\nu}\equiv-\tb_{\mu\nu}+\tilde\calT_{\mu\nu}=0.
\end{eqnarray}
We introduce $\calT^\lambda$ and $\tilde\calT^\lambda$ according to 
\eqno(3,3), so that
\begin{eqnarray}
&&\calT^\lambda\equiv -\tb^\lambda+\tilde\calT^\lambda=0.
\end{eqnarray}
We note that \eqno(3,18) is rewritten as
\begin{eqnarray}
&&\xi_{\lambda\mu\nu}\partial^\mu\calT^\nu=0.
\end{eqnarray}
\par
For calculating Wightman functions, we should not use \eqno(4,7), but write
\begin{eqnarray}
&&\tilde\calT^+\equiv -i(2\bar c^+\partial_+c^++\partial_+\bar c^+\cdot c^+)
                +\partial_+\phi_M\cdot\partial_+\phi^M.
\end{eqnarray}
Then from the formulae given in Sec.\ 5, we obtain
\begin{eqnarray}
&&\Wightman{\tb^+(x_1)\tb^+(x_2)}=0,\\
&&\Wightman{\tb^+(x_1)\tilde\calT^+(x_2)}=(D-26)\Phi^{++},\\
&&\Wightman{\tilde\calT^+(x_1)\tilde\calT^+(x_2)}=(D-26)\Phi^{++},\\
\noalign{\noindent where}
&&\qquad \Phi^{++}\equiv {1\over 2(2\pi)^2}\cdot 
                        {1\over (x_1{}^+-x_2{}^+-i0)^4}\not=0.
\end{eqnarray}
We thus see that
\begin{eqnarray}
&&\Wightman{\tb^+(x_1)\calT^+(x_2)}=(D-26)\Phi^{++},\\
&&\Wightman{\calT^+(x_1)\calT^+(x_2)}=-(D-26)\Phi^{++},
\end{eqnarray}
that is, we encounter the field-equation anomaly because \eqno(6,2) is violated
at the level of Wightman functions.
Of course, the once-differentiated equation \eqno(6,3) is not broken at all.
\par
The field-equation anomaly has arisen because $\Wightman{\tb^+\tb^+}$ is
not equal to $\Wightman{\tb^+\tilde\calT^+}
=\Wightman{\tilde\calT^+\tilde\calT^+}$.
From \eqno(4,11), it is impossible to make the former equal to
$(D-26)\Phi^{++}$ even if we dare to violate the BRS invariance (or FP-ghost
number) so as to have $\Wightman{\tb^+}\not=0$.  Thus the appearance of the 
field-equation anomaly is unavoidable.
\par
We can extend the consideration given in \eqno(6,5)-\eqno(6,10) to the 
3-point functions.  We find that
\begin{eqnarray}
&&\Wightman{\tb^+(x_1)\tb^+(x_2)\tb^+(x_3)}=0, \\
&&\Wightman{\tb^+(x_1)\tb^+(x_2)\tilde\calT^+(x_3)}
 =\Wightman{\tb^+(x_1)\tilde\calT^+(x_2)\tilde\calT^+(x_3)} 
 =\Wightman{\tilde\calT^+(x_1)\tilde\calT^+(x_2)\tilde\calT^+(x_3)} \nonumber\\
&&\quad
 ={-1\over(2\pi)^3}\cdot 
   {D-26\over (x_1{}^+-x_2{}^+-i0)^2(x_2{}^+-x_3{}^+-i0)^2
              (x_1{}^+-x_3{}^+-i0)^2}.
\end{eqnarray}
The appearance of $D-26$ is quite stable: For any linear or quadratic local 
operators $F_j{}^+(x)$, we see that $\Wightman{F_1{}^+(x_1)\calT^+(x_2)}$ and 
$\Wightman{F_1{}^+(x_1)F_2{}^+(x_2)\calT^+(x_3)}$ are either zero or 
proportional to $D-26$.
\par
In spite of the presence of the field-equation anomaly, we can define various
symmetry generators so as to be free of its trouble.
We here present the anomaly-free definitions of the translation generator
$P_\nu$, the BRS generator $Q_b$ and the FP-ghost number generator $Q_c$.
\par
The Noether currents of translation, BRS and FP-ghost number are
\begin{eqnarray}
&&J_\nu{}^\mu=-{i\over\sqrt2}\xi^\mu{}_{\lambda\rho}\bar c^\lambda
                \partial_\nu c^\rho
              +\partial_\nu\phi_M\cdot\partial^\mu\phi^M
              -{1\over2}\delta_\nu{}^\mu\partial_\sigma\phi_M\cdot
               \partial^\sigma\phi^M, \\
&&j_b{}^\mu=-{i\over\sqrt2}\xi^\mu{}_{\lambda\rho}\bar c^\lambda c^\sigma
             \partial_\sigma c^\rho
            -c^\sigma\partial_\sigma\phi_M\cdot\partial^\mu\phi^M
            +{1\over2}c^\mu\partial_\sigma\phi_M\cdot\partial^\sigma\phi^M, \\
&&j_c{}^\mu=-{i\over\sqrt2}\xi^\mu{}_{\lambda\rho}\bar c^\lambda c^\rho,
\end{eqnarray}
respectively.  In the light-cone representation, they reduce to
\begin{eqnarray}
&&J_+{}^-=-i\bar c^+\partial_+c^++\partial_+\phi_M\cdot\partial_+\phi^M, \\
&&j_b{}^-=-\bar c^+c^+\partial_+c^+-c^+\partial_+\phi_M\cdot\partial_+\phi^M,\\
&&j_c{}^-=-i\bar c^+ c^+.
\end{eqnarray}
Noting \eqno(6,4), we rewrite \eqno(6,16) and \eqno(6,17) as
\begin{eqnarray}
&&J_+{}^-=\tb^++\calT^++i\partial_+(\bar c^+ c^+),\\
&&j_b{}^-=-\tb^+c^+ +i\bar c^+c^+\partial_+c^+-\calT^+c^+,
\end{eqnarray}
respectively.
\par
Since the terms involving $\calT^+$ suffer from the field-equation anomaly,
we drop them.  Thus the anomaly-free generators are defined by
\begin{eqnarray}
&&P_\pm\equiv \int dx^\pm \tb^\pm, \\
&&Q_b\equiv \int dx^+(-\tb^+c^++i\bar c^+c^+\partial_+c^+)
           +\int dx^-(-\tb^-c^-+i\bar c^-c^-\partial_-c^-), \\
&&iQ_c\equiv\int dx^+\bar c^+c^+ +  \int dx^-\bar c^- c^-.
\end{eqnarray}
\vskip50pt
%
%%%%%%%%%%%%%%%%%%%%%%%%%%%%%  Section 7  %%%%%%%%%%%%%%%%%%%%%%%%%%
%
\Sec{Perturbative approach to the conformal anomaly}
In this section, we review the perturbative approach to the conformal anomaly
in order to compare it with our exact results.  Since $\lagGF$ contains no
differentiation, the B-field $\tb_{\mu\nu}$ is nonpropagating in perturbation
theory, and therefore it is customary to discard it.
Then the conformal-gauge two-dimensional quantum gravity reduces to a fee
field theory.  Nevertheless, one wishes to encounter the conformal anomaly.
The procedure for this\cite{Baulieu-Bilal} is as follows.
\par
From \eqno(2,12) with \eqno(2,13), \eqno(2,14) and \eqno(2,6), the free 
Lagrangian density is given by
\begin{eqnarray}
\lag^{(0)}&=&{1\over2}\eta^{\mu\nu}\partial_\mu\phi_M\cdot\partial_\nu\phi^M
             -{1\over2}\tb_{\mu\nu}h^{\mu\nu} \nonumber \\
          & &-{1\over2}i\bar c_{\mu\nu}(\eta^{\mu\sigma}\partial_\sigma c^\nu
               +\eta^{\nu\sigma}\partial_\sigma c^\mu
               -\eta^{\mu\nu}\partial_\sigma c^\sigma).
\end{eqnarray}
One then introduces a background field $\hg^{\mu\nu}$ and makes \eqno(7,1)
background-covariant by replacing $\eta^{\mu\nu}$ by $\thg^{\mu\nu}$
and $\partial_\mu$ by background-covariant differentiation $\hat\nabla_\mu$.
In this way, one obtains
\begin{eqnarray}
\hat\lag&=&{1\over2}\thg^{\mu\nu}\partial_\mu\phi_M\cdot
                                           \partial_\nu\phi^M
          -{1\over2}\tb_{\mu\nu}h^{\mu\nu} \nonumber \\
& &-{1\over2}i\bar c_{\mu\nu}[\thg^{\mu\sigma}\hnabla_\sigma c^\nu
             +\thg^{\nu\sigma}\hnabla_\sigma c^\mu
             -\hnabla_\sigma(\thg^{\mu\nu}c^\sigma)].
\end{eqnarray}
Here it is very interesting to note that the quantity in the square bracket of
\eqno(7,2) can be rewritten as
\begin{eqnarray}
&&\thg^{\mu\sigma}\partial_\sigma c^\nu
  +\thg^{\nu\sigma}\partial_\sigma c^\mu
  -\partial_\sigma(\thg^{\mu\nu}c^\sigma) 
\end{eqnarray}
identically.
\par
The ``energy-momentum tensor'' $T_{\mu\nu}$ is defined by
\begin{eqnarray}
T_{\mu\nu}&\equiv&2{\delta\over\delta\hg^{\mu\nu}}
            \int d^2x\,\hat\lag\bigg|_{\hg^{\mu\nu}=\eta^{\mu\nu}} 
           \nonumber\\
&=&-i[\bar c_{\mu\sigma}\partial_\nu c^\sigma
     +\bar c_{\nu\sigma}\partial_\mu c^\sigma
     +\partial_\sigma\bar c_{\mu\nu}\cdot c^\sigma
     -\eta_{\mu\nu}\bar c_{\sigma\tau}\partial^\sigma c^\tau] \nonumber \\
& &+\partial_\mu\phi_M\cdot\partial_\nu\phi^M
   -{1\over2}\eta_{\mu\nu}\partial_\sigma\phi_M\cdot\partial^\sigma\phi^M.
\end{eqnarray}
In contrast with the covariant-gauge case, $T_{\mu\nu}$ is traceless.
One calculates the 2-point functions of $T_{\mu\nu}$ by using Feynman 
propagators
\begin{eqnarray}
&&\wightman{\bar c_{\mu\nu}(x) c^\lambda(y)}
 =-i(\delta_\mu{}^\lambda\partial_\nu
    +\delta_\nu{}^\lambda\partial_\mu
    -\eta_{\mu\nu}\partial^\lambda)\Df(x-y), \\
&&\wightman{\phi_M(x)\phi^N(y)}
 =\delta_M{}^N\Df(x-y).
\end{eqnarray}
One obtains
\begin{eqnarray}
&&\wightman{T_{\mu\nu}(x)T_{\lambda\rho}(y)}
 =(D-26)\Phi_{\mu\nu\lambda\rho}(x-y)\ +\ \hbox{local terms},
\end{eqnarray}
where the Fourier transform of $\Phi_{\mu\nu\lambda\rho}$ is proportional to 
$p_\mu p_\nu p_\lambda p_\rho / (p^2+i0)$.
The nonlocal term of \eqno(7,7) is called the conformal anomaly.
\par
Owing to the identity noted in \eqno(7,3), we have a remarkable 
equality\cite{AN1}
\begin{eqnarray}
&&T_{\mu\nu}=\tilde\calT_{\mu\nu},
\end{eqnarray}
where $\tilde\calT_{\mu\nu}$ is the quantity defined in \eqno(6,1).
Hence \eqno(7,7) is essentially nothing but \eqno(6,7).
It should be noted that $\tilde\calT_{\mu\nu}$ is {\it different\/} from 
$\calT_{\mu\nu}$; the latter only is the sensible quantity in the exact
theory.  It is an accidental coincidence that both \eqno(6,7) and \eqno(6,10)
are proportional to $D-26$.
\par
The inadequacy of the perturbative approach can clearly be seen by considering
other anomalies.  The FP-ghost number current anomaly\cite{Kraemmer-Rebhan} is
obtained by calculating the nonlocal term of 
$\wightman{j_c{}^\mu(x)T_{\lambda\rho}(y)}$, which is found to be nonvanishing.
It should be noted, however, that
\begin{eqnarray}
\wightman{j_c{}^\mu(x)T_{\lambda\rho}(y)}
&=&\wightman{j_c{}^\mu(x)\tilde\calT_{\lambda\rho}(y)} \nonumber \\
&\not=&\wightman{j_c{}^\mu(x)\calT_{\lambda\rho}(y)}=0.
\end{eqnarray}
Indeed, explicit calculation shows that
\begin{eqnarray}
\Wightman{j_c{}^-(x)\calT^{+}(y)}
&=&
-\Wightman{j_c{}^-(x)\tb^+(y)}+\Wightman{j_c{}^-(x)\tilde\calT^{+}(y)} \nonumber\\
&=&
-{3\over(2\pi)^2}\cdot{1\over(x^+-y^+-i0)^3}
     +{3\over(2\pi)^2}\cdot{1\over(x^+-y^+-i0)^3}=0. \hspace*{40pt}
\end{eqnarray}
The {\it non-existence\/} of the FP-ghost number current anomaly is quite
consistent with the exact solution given in Sec.\ 5.
Thus perturbative approach is quite misleading.
\vfill\eject
%
%%%%%%%%%%%%%%%%%%%%%%%%%%%%%  Section 8  %%%%%%%%%%%%%%%%%%%%%%%%%%%%
%
\Sec{Discussion}
In his paper entitled ``Quantum gravity in two dimensions'',\cite{Polyakov}
Polyakov wrote ``The most simple form this formula (i.e., Polyakov's nonlocal
action) takes is in the conformal gauge, where $g_{ab}=e^\vphi \delta_{ab}$
where it becomes a free field action.  Unfortunately this simplicity is an 
illusion.'' And he adopted the light-cone gauge.
Even if we employ the BRS quantization, the conformal-gauge two-dimensional
quantum gravity becomes a free field theory if the B-field is discarded.
Nevertheless, the critical dimension $D=26$ is obtained in this model.  
In the present paper, we have clarified why such a paradoxical phenomenon 
happens. Our conclusion is as follows.
\par
{\it The field equation (6.1) for the B-field suffers from the field-equation
anomaly.} That is, \eqno(6,1) is valid at the level of operator algebra, but
it is violated {\it modulo\/} \eqno(6,3) at the level of the representation
in terms of state vectors.   Therefore, if one discards the B-field under the
understanding that \eqno(6,1) is nothing more than a definition of the B-field,
one necessarily misses the existence of the field-equation anomaly.
We have found that \eqno(6,1) is {\it not\/} a mere defining equation: It is 
not a trivial statement to set up an equality between a fundamental field, 
which is a BRS transform of the FP-antighost, and a certain composite operator,
which has no linear term of fundamental fields.  The anomalous behaviors of
the conformal-gauge two-dimensional quantum gravity are the consequence of the 
field-equation anomaly for \eqno(6,1).  The reason why people have never been
aware of this fact is that they always adopted the path-integral type approach
so that they could not clearly distinguish the operator level and the 
representation level. Indeed, for instance, Fujikawa\cite{Fujikawa} eliminates
the B-field at the first step, so that the relevance of (6.1) to the anomaly
is not explicitly recognized in his calculation.
\par
In the conformal gauge, the field-equation anomaly seems to be always 
proportional to $D-26$.  This is a very special situation of the conformal
gauge.  What is obtained from perturbative approach, however, is not identical
with this ``$D-26$''.  That is, the formula \eqno(7,7) which gives the 
``conformal anomaly'' precisely proportional to $D-26$ is, owing to \eqno(7,8),
nothing but \eqno(6,7) {\it but not (6.10)\/}. 
As discussed in our previous papers,\cite{AN1,AN8} the nonlocal term called
``conformal anomaly'' is {\it produced\/} by the operation 
$\delta/\delta \hat g^{\mu\nu}$.  This fact is more clearly seen in the 
FP-ghost number current anomaly: While the exact solution given in Sec.\ 5 is
completely consistent with the FP-ghost number conservation, the perturbative
approach implies the existence of its anomaly.\cite{Kraemmer-Rebhan}
This paradoxical result can be explained by recognizing that the FP-ghost
number current anomaly has been {\it produced\/} by the operation
$\delta/\delta \hat g^{\mu\nu}$,\cite{AN1} that is, it is a consequence of 
using the artificial quantity $T_{\mu\nu}$.
\par
Anyway, the conformal-gauge two-dimensional quantum gravity implies 
the existence of the filed-equation anomaly proportional to $D-26$.  
Thus at the critical dimension $D=26$ the field-equation anomaly is absent
in the conformal gauge.
This fact is quite natural because the conformal-gauge two-dimensional 
quantum gravity, which is definable only in the {\it strictly\/} 
two-dimensional case, is a theory lying in between the string theory of 
finite length and the covariant (de Donder)-gauge two-dimensional quantum
gravity.
\par
Finally, we note that in the conformal gauge perturbative approach is stable
in contrast with the case of covariant gauge.  This is due to the fact that
the gauge-fixing Lagrangian density in the conformal gauge contains no linear
term in the sense of perturbation theory. 
In the de Donder-gauge case,\cite{AN1}  such a redefinition of the B-field as
\begin{eqnarray}
&&\tb_\rho=b_\rho+ic^\sigma\partial_\sigma\bar c_\rho
\end{eqnarray}
changes the quadratic part of the FP-ghost Lagrangian density, 
which contributes to the ``conformal anomaly''.\cite{AN9}
In the conformal-gauge case, \eqno(8,1) brings no change to the quadratic 
part.
\par
The B-field $b_\rho$ appearing in the right-hand side of \eqno(8,1) is the 
{\it intrinsic\/} B-field, which is regarded as the primary B-field more
natural that $\tb_\rho$.\cite{AN1}
It should be noted, however, that in contrast with the de Donder-gauge case,
it is impossible to define the intrinsic B-field in the conformal-gauge case
because in the latter $\sqrt{-g}$ is not available so that we cannot define
the action invariant under the intrinsic BRS transformation ($ib_\rho$ is
the intrinsic BRS transform of the FP antighost).  All such circumstances are
quite consistent with the stability of $D-26$ in the conformal gauge.
\vfill\eject
%
%%%%%%%%%%%%%%%%%%%%%%%%%  References  %%%%%%%%%%%%%%%%%%%%%%%%%
%

\end{document}